\documentclass[10pt,journal,compsoc]{IEEEtran}
\ifCLASSOPTIONcompsoc
  \usepackage[nocompress]{cite}
\else
  \usepackage{cite}
\fi
\ifCLASSINFOpdf
  \usepackage[pdftex]{graphicx}
\else
\fi
\usepackage{amsmath}
\usepackage{amsfonts,amssymb}
\usepackage{algorithmic}
\usepackage{algorithm}
\usepackage[caption=false,font=footnotesize,labelfont=sf,textfont=sf]{subfig}

\usepackage{stfloats}
\begin{document}
%
\title{A BDI Agent-Based Task Scheduling Framework for Cloud Computing}
%
%
%
%

\author{Yikun Yang,~\IEEEmembership{Student~Member,~IEEE,}
        Fenghui Ren,~\IEEEmembership{Fellow,~IEEE,}
        and~Minjie~Zhang,~\IEEEmembership{Fellow,~IEEE}
}

\IEEEtitleabstractindextext{%
\begin{abstract}
Cloud computing is an attractive technology for providing computing resources over the Internet. Task scheduling is a critical issue in cloud computing, where an efficient task scheduling method can improve overall cloud performance. Since cloud computing is a large-scale and geographically distributed environment, traditional scheduling methods that allocate resources in a centralized manner are ineffective. Besides, traditional methods are difficult to make rational decisions timely when the external environment changes. This paper proposes a decentralized BDI (belief-desire-intention) agent-based scheduling framework for cloud computing. BDI agents have advantages in modelling dynamic environments because BDI agents can update their beliefs, change desires, and trigger behaviours based on environmental changes. Besides, to avoid communication stuck caused by environmental uncertainties, the asynchronous communication mode with a notify listener is employed. The proposed framework covers both the task scheduling and rescheduling stages with the consideration of uncertain events that can interrupt task executions. Two agent-based algorithms are proposed to implement the task scheduling and rescheduling processes, and a novel recommendation mechanism is presented in the scheduling stage to reduce the impact of information synchronization delays. The proposed framework is implemented by JADEX and tested on CloudSim. The experimental results show that our framework can minimize the task makespan, balance the resource utilization in a large-scale environment, and maximize the task success rate when uncertain events occur.
\end{abstract}

\begin{IEEEkeywords}
Cloud Computing, Belief-Desire-Intention, Multi-Agent System, Task Scheduling.
\end{IEEEkeywords}}

\maketitle

\IEEEdisplaynontitleabstractindextext

%
\IEEEpeerreviewmaketitle

\IEEEraisesectionheading{\section{Introduction}\label{sec:introduction}}

%
%
%
%
\IEEEPARstart{W}{ith} the rapid advancement of communication and Internet of Things (IoT) technologies, vast amounts of data concurrency and exchange over the Internet \cite{arunarani2019task,fellir2020multi}. Local computing resources, such as PC and local servers, are gradually unable to satisfy growing user demands. Cloud computing provides shared and powerful computing resources to cloud users commercially and elastically, where cloud providers and users set the Service-Level Agreement (SLA) to characterize the Quality of Service (QoS) \cite{gupta2017load,zhong2016virtual}. In cloud computing, task scheduling is critical in developing the QoS. Task scheduling refers to mapping available computing resources to cloud users and achieving scheduling objectives, such as improving the system throughput.

Unlike the centralized resource allocation problems in traditional manufacturing scenarios, cloud services are delivered via the Internet, so the whole environment is large-scale and geographically distributed. Three challenging issues need to be considered to address the task scheduling problem in cloud computing. (1) Since cloud users are distributed worldwide and can request computing resources at any time, it is difficult to collect and process global task information centrally. Hence, traditional scheduling methods that require global information on tasks and resources are difficult to be applied in cloud computing. 
(2) The billions of cloud users with various task requirements make task scheduling introduce a massive amount of computation. The large-scale user requests could increase the system load and latency, as well as the decision-making time on scheduling, so as to impact the scheduling optimization. (3) The operations of distributed users and the availability of computing resources are unpredictable. When uncertain events invalidate the current schedule, scheduling methods are required to reallocate available resources for tasks to ensure task executions. However, most existing scheduling methods for cloud computing lack the rescheduling mechanism for maximizing task success rate. 

Considering the three challenging issues mentioned above, this paper proposes a decentralized BDI (Belief-Desire-Intention) agent-based framework for cloud task scheduling. Three contributions of the framework are shown as follows. (1) The framework employs BDI agents with the asynchronous communication mode to implement scheduling and rescheduling processes. BDI agents have advantages in modelling complex distributed scheduling environments, which can distribute the scheduling complexity to individual agents through parallel computations. The asynchronous communication mode can reduce the risk of agents being stuck when they fail to receive expected responses to improve the interaction robustness. (2) This paper proposes a novel agent-based algorithm named \textit{Asynchronous Recommendation Algorithm} (ARA) to minimize the conflicts among decentralized agents caused by their self-interests. Besides, the ARA can help to balance two different scheduling objectives, (i) minimizing task makespan and (ii) balancing resource utilization. (3) This paper proposes an agent-based solution to maximize the task success rate when uncertain events occur. The proposed rescheduling algorithm applies the BDI model to detect uncertain events as \textit{belief} changes and automatically trigger corresponding \textit{desires} and \textit{intentions} to reallocate available resources for affected tasks. 

The rest of this paper is organized as follows. In Section \ref{related_work}, related work is given. In Section \ref{background_and_statement}, the considered cloud computing environment is defined and the research problem is described. In Section \ref{framework_design}, the proposed agent-based framework and two agent-based algorithms are presented. In Section \ref{experiment}, experiments and results are discussed and the conclusion and future work are given in Section \ref{conclusion}.

\section{Related Work}\label{related_work}

This section introduces related works in two aspects, (1) the task scheduling problem in cloud computing and (2) multiagent-based task scheduling methods.

\subsection{Task Scheduling in Cloud Computing}

In recent years, task scheduling has been treated as a critical challenge in cloud computing \cite{arunarani2019task,ibrahim2021task}.
In the literature, non-agent-based task scheduling methods for cloud computing are mainly classified into \textbf{static scheduling methods}, and \textbf{dynamic scheduling methods} \cite{mathew2014study}.

\textit{Static scheduling methods} assume that all tasks arrive at the cloud system simultaneously and the global information is known in advance. Static scheduling methods mainly focus on the optimization of scheduling results. Heuristic algorithms Minimum Execution Time (MET) and Minimum Completion Time (MCT) are two classic static scheduling methods \cite{nagadevi2013survey}. The MET maps tasks and resources based on the shortest execution time, while the MCT maps tasks and resources based on the earliest completion time. 
Besides, heuristic algorithms, such as the Genetic Algorithm \cite{yiqiu2019cloud} and Simulated Annealing Algorithm \cite{gabi2018hybrid}, can help to optimize the scheduling results in static environments. For example, in \cite{yiqiu2019cloud}, the Genetic Algorithm randomly generates the initial population to represent the mapping between tasks and resources. Then it operates the population through selection, crossover, and mutation to optimize the schedule.

\textit{Dynamic scheduling methods} assume that tasks arrive at the cloud system during the runtime, and the global information is not known in advance. In this case, the scheduler maps tasks on resources after they arrive. 
First Come First Serve (FCFS) \cite{ramkumar2019preserving}, Round-Robin \cite{sanaj2020enhanced}, Min-Min \cite{derakhshan2018optimization}, and Max-Min \cite{bhoi2013enhanced}, are widely discussed in the literature as dynamic scheduling methods. For example, the FCFS executes the waiting tasks in order of their arrival, and the Round-Robin allocates time slots of resources to tasks in equal portions and in circular order. Min-Min and Max-Min algorithms schedule tasks on resources that provide the earliest completion time. The only difference is that Min-Min selects the shortest task first, while Max-Min selects the longest task first. 

In addition to the time-oriented scheduling methods mentioned above, factors such as the energy consumption \cite{ben2019efficient,ding2020q}, service costs \cite{jiang2019pivot,jiang2020cloud}, fault-tolerant \cite{han2018fault,lee2019adaptive,li2021real}, and resource load balancing \cite{hussain2018ralba,mishra2020load} have been deeply studied in the literature which also contribute to the QoS of cloud computing.

The main limitation of non-agent-based methods is the centralized decision-making structure, where the central scheduler makes all scheduling decisions. The entities in cloud computing are geographically distributed and the scale is quite large. Hence, the scheduling information is hard to collect centrally, and the scheduler is challenging to generate optimal schedules within a reasonable time for large-scale tasks.

\subsection{Multiagent-based Task Scheduling}

Multiagent-based approaches with a decentralized structure have demonstrated advantages in modelling large-scale distributed systems \cite{bhutta2020loop,wang2020large,yang2021agent}. 
In the literature, several multiagent-based task scheduling approaches were proposed \cite{d2021designing,zhu2015angel}.
For instance, D'Aniello et al. \cite{d2021designing} proposed a multi-agent system for managing and monitoring services in the cloud manufacturing system and minimizing the impact of both dynamic task arrival and resource downtime. In their design, three types of agents (Task Agent, Master Agent, and Printer Agent) were proposed to monitor the environment states and to negotiate with each other for task scheduling. Specifically, the Master Agent receives the task batch from Task Agents and the availability information from Printer Agents, and then performs the scheduling algorithm for each task batch. In \cite{zhu2015angel}, Zhu et al. proposed an agent-based scheduling mechanism with a novel bidirectional announcement bidding mechanism to allocate real-time tasks in cloud computing. This approach takes the elasticity of cloud computing into consideration, and new virtual machines can be added to the cloud dynamically. 

In the literature, several studies discussed the uncertainty in cloud computing and the necessity of rescheduling \cite{kabir2021uncertainty,tchernykh2019towards}. However, agent-based rescheduling methods were rarely presented. Besides, most of the current agent-based scheduling methods are based on negotiations and cooperative decision-making between intelligent agents. Two potential limitations make current agent-based approaches inflexible and not robust in large-scale cloud computing environments. (1) Most decision-making models in current approaches are rule-based, which makes the agent unable to cope with unknown situations, so as to decrease the flexibility. (2) Most agent interactions are based on the synchronous communication mode (i.e., FIPA interaction protocols), where agents easily get stuck once they cannot receive replies as expected, thereby reducing the robustness.

To improve the scheduling flexibility and robustness, this paper applies the BDI agent (belief-desire-intention decision-making model \cite{de2020bdi,nunes2011bdi4jade}) with the asynchronous communication mode to implement the scheduling framework for cloud computing. Two novel agent-based algorithms are proposed to coordinate the behaviours of agents in the scheduling and rescheduling processes, respectively.

\section{Problem Description} \label{background_and_statement}

This section describes the cloud scheduling environments and the scheduling and rescheduling objectives.  

\subsection{Scheduling Environments in Cloud Computing}

\begin{figure}[!t]
\centering
\includegraphics[width=3.5in]{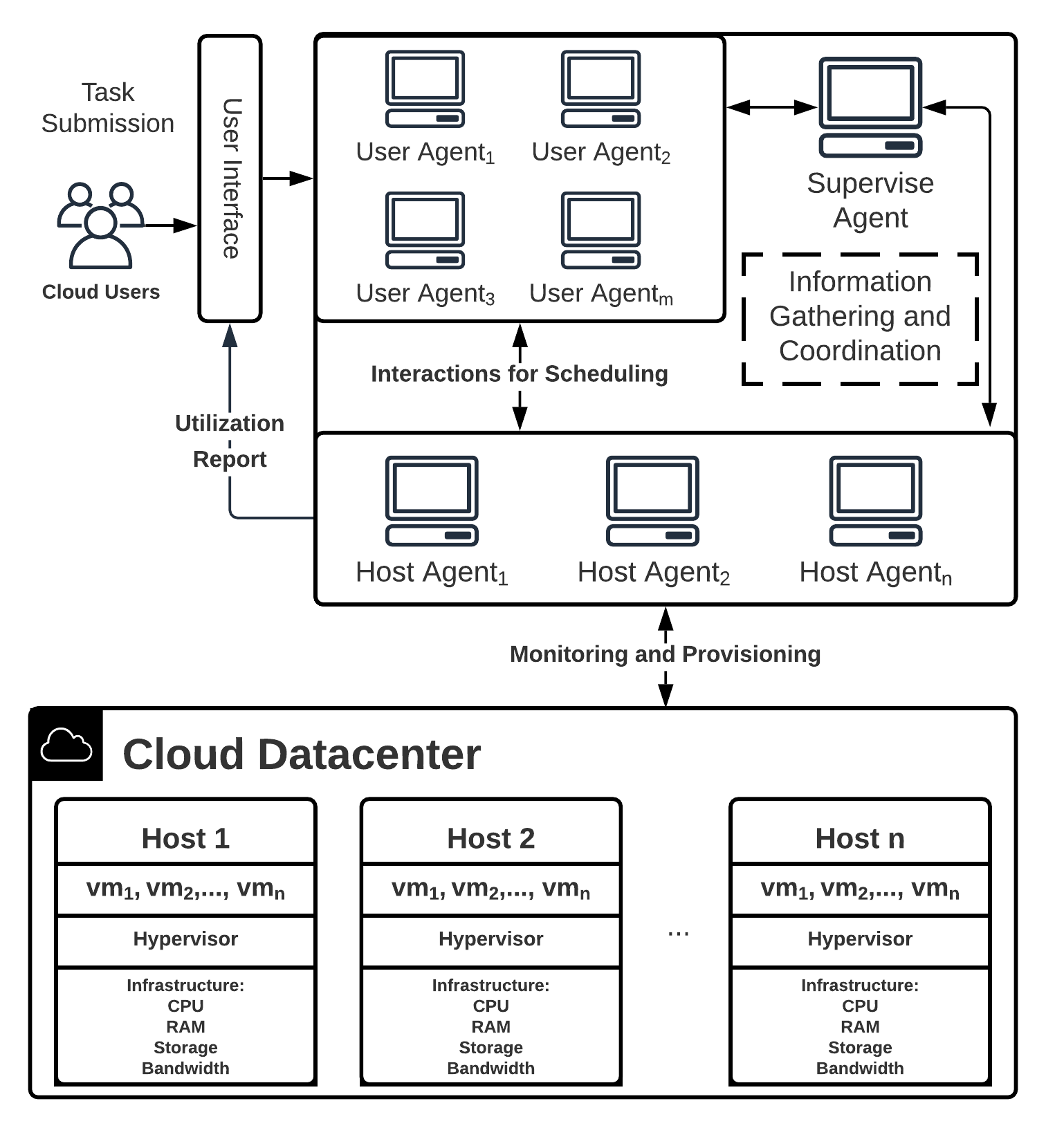}
\caption{Framework of the considered cloud computing environment and proposed MAS.}
\label{architecture_diagram}
\end{figure}

This paper considers the IaaS (Infrastructure-as-a-Service) based cloud computing which delivers storage and computing resources to cloud users over the Internet in a pay-as-you-go manner. 
As shown in Fig \ref{architecture_diagram}, tasks are submitted by cloud users and executed by virtual machines in the cloud datacenter. This paper focuses on dynamic task scheduling with \textit{hard-deadline}, where new cloud users with unknown tasks can be added to the cloud environment at runtime, and the task deadline is strict.

A cloud datacenter $DC = \left\lbrace H_{1},..., H_{i} \right\rbrace$ administers a set of hosts, where $H_{i}$ represents the i-th host in $DC$. The infrastructure characteristics (i.e., CPU, RAM, storage, and network bandwidth) of a host are fixed, which characterize the computing capabilities of virtual machines (VMs) governed by the host. 

A host $H_{i} = \left\lbrace vm_{i}^{1},...,vm_{i}^{k} \right\rbrace$ includes a set of virtual machines, where $vm_{i}^{k}$ represents the k-th virtual machine in the i-th host. 
The k-th virtual machine of the host $H_{i}$ is defined as $vm_{i}^{k} = \left\lbrace cpu_{i}^{k}, ram_{i}^{k}, sc_{i}^{k}, bw_{i}^{k}, at_{i}^{k}(\tau) \right\rbrace$, where $cpu_{i}^{k}$ denotes the MIPS of CPU, $ram_{i}^{k}$ denotes the RAM capacity, $sc_{i}^{k}$ denotes the storage capacity, $bw_{i}^{k}$ denotes the network bandwidth, and the $at_{i}^{k}(\tau)$ denotes the next available time at time $\tau$. 

Cloud users $U = \left\lbrace u_{1},..., u_{n} \right\rbrace$ submit the task requirements and the specified deadline through the User Interface (UI). Requests of the n-th user can be represented as $u_{n} = \left[T_{n}, D_{n} \right]$, where $T_{n}=\left\lbrace t_{n}^{1},...,t_{n}^{p}\right\rbrace$ includes a set of independent tasks, and $D_{n}$ denotes the hard-deadline of $u_{n}$. The p-th task of the $u_{n}$ is defined as $t_{n}^{p}=\left\lbrace wl_{n}^{p}, ram_{n}^{p}, sc_{n}^{p}, bw_{n}^{p} \right\rbrace$, where $wl_{n}^{p}$ gives the workload of the task (given in MI), $ram_{n}^{p}$ represents the RAM requirement of the task, $sc_{n}^{p}$ denotes the task storage requirement, and $bw_{n}^{p}$ denotes the task bandwidth requirement.

\subsection{Scheduling Objectives}

Multiple scheduling objectives have been studied for cloud computing \cite{hosseinzadeh2020multi}. 
This paper mainly considers two scheduling objectives and one rescheduling objective. Two scheduling objectives are (1) to minimize the makespan and (2) to balance the resource utilization. The rescheduling objective is to maximize the task success rate. Three objective functions are formulated as follows:

1. Let $C_{n}$ denote the task completion time of user $u_{n}$ and $C_{max}= \max\limits_{u_{n} \in U} \left\lbrace C_{n} \right\rbrace$ denote the maximum completion time of all users, the first scheduling objective is the minimization of $C_{max}$, i.e., $\min C_{max}$.

2. Let $\mu_{i}= AT_{i}/RC_{i}$ denote the resource utilization of a virtual machine $vm_{i}$, where $AT_{i}$ denotes the allocated time for $vm_{i}$, and $RC_{i}$ denotes the resource capacity. Let $V(\mu)=\sum_{i=1}^{N}(\mu_{i} - \overline{\mu} )^{2}/NV$ denote the utilization variance, where $NV$ denotes the total number of virtual machines, $\overline{\mu}$ denotes the average resource utilization. The second scheduling objective is to balance the resource utilization, i.e., $\min V(\mu)$.

3. Let $SR= SN/NT$ denote the task success rate, where $SN$ indicates the number of successful tasks, and $NT$ is the total number of tasks submitted to the system. The rescheduling objective is to maximize the task success rate, i.e., $max SR$.

\section{Agent-Based Framework Design} \label{framework_design}

This section introduces the architecture of the proposed framework and describes the behaviours of agents during the scheduling and rescheduling stages. 

\subsection{Framework Architecture}

As Fig. \ref{architecture_diagram} shows, the proposed agent-based framework includes three agent modules, (1) an \textit{user agent module}, (2) a \textit{host agent module}, and (3) a \textit{supervise agent module}. Three types of agents are implemented by the BDI decision-making model and equipped with the asynchronous communication mode.

\begin{figure}[h!]
\centering
\includegraphics[width=3.5in]{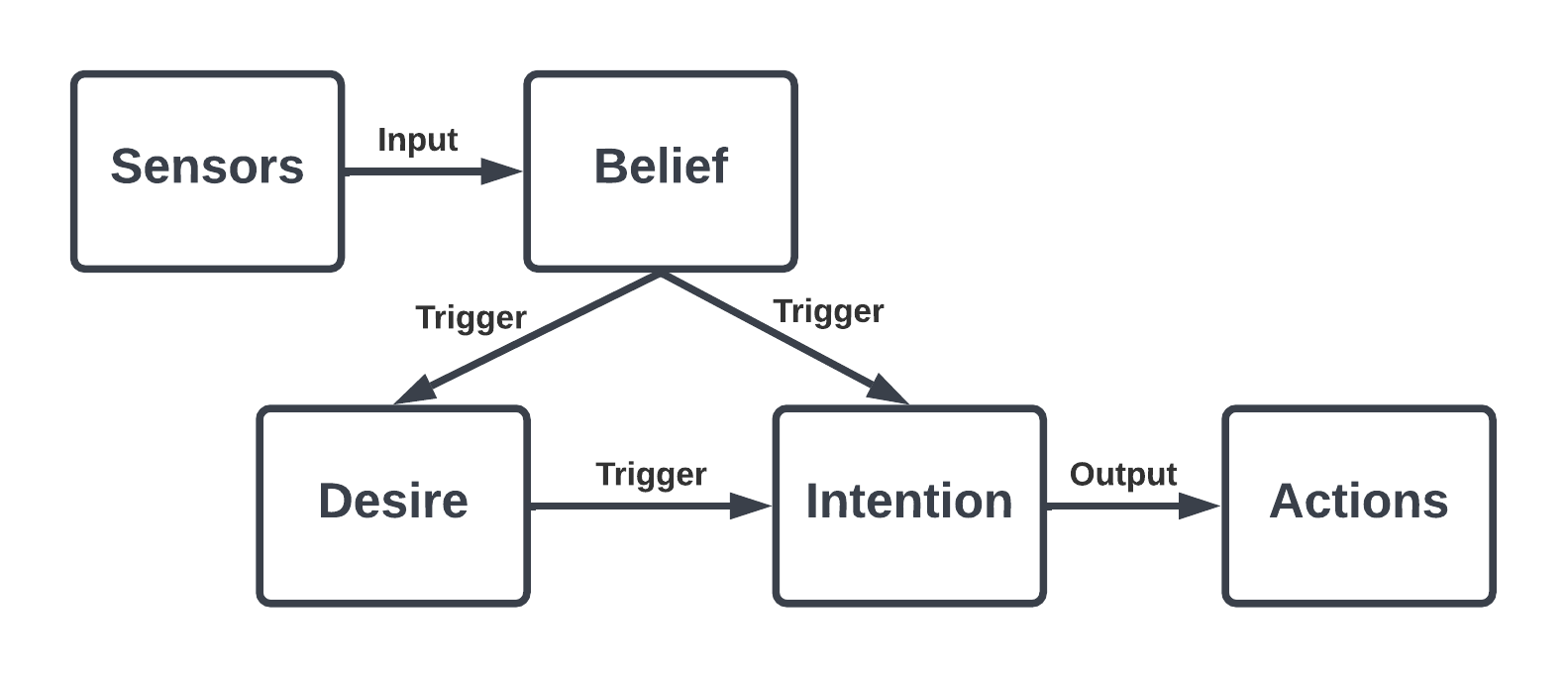}
\caption{Belief-Desire-Intention Model.}
\label{bdi_model}
\end{figure}

Fig. \ref{bdi_model} shows the structure of the BDI model. Each BDI agent is defined by its \textbf{beliefs}, \textbf{desires}, and \textbf{intentions} \cite{nunes2011bdi4jade}. Beliefs represent the internal and external knowledge states of the agent. Through interactions with the external environment via sensors, agent beliefs can change during the process, and the change can trigger new desires and intentions. Desires represent the objectives or goals the agent aims to achieve in different states, and intentions represent multiple plans the agent can take to achieve desires.

\begin{figure*}[t!]
\centering
\subfloat[Synchronous Mode]{\includegraphics[height=1.8in]{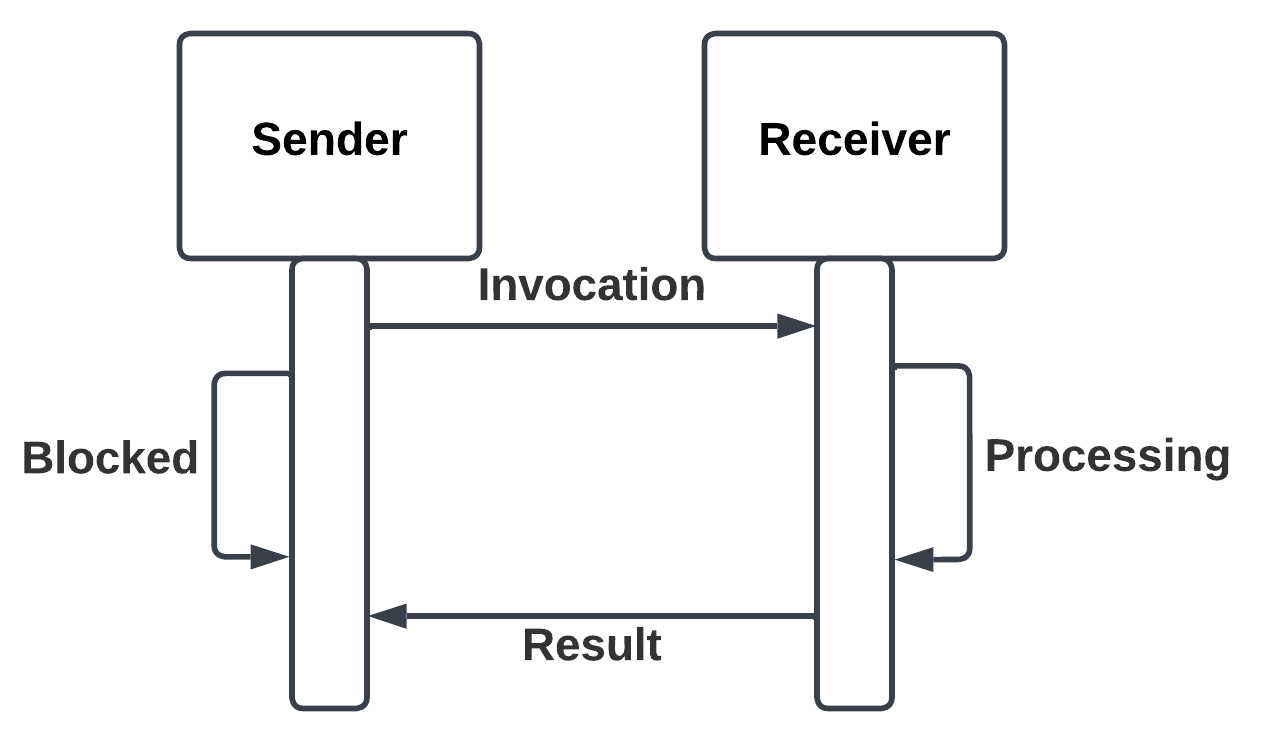}%
\label{syn_call}}
\hfil
\subfloat[Asynchronous Mode]{\includegraphics[height=1.8in]{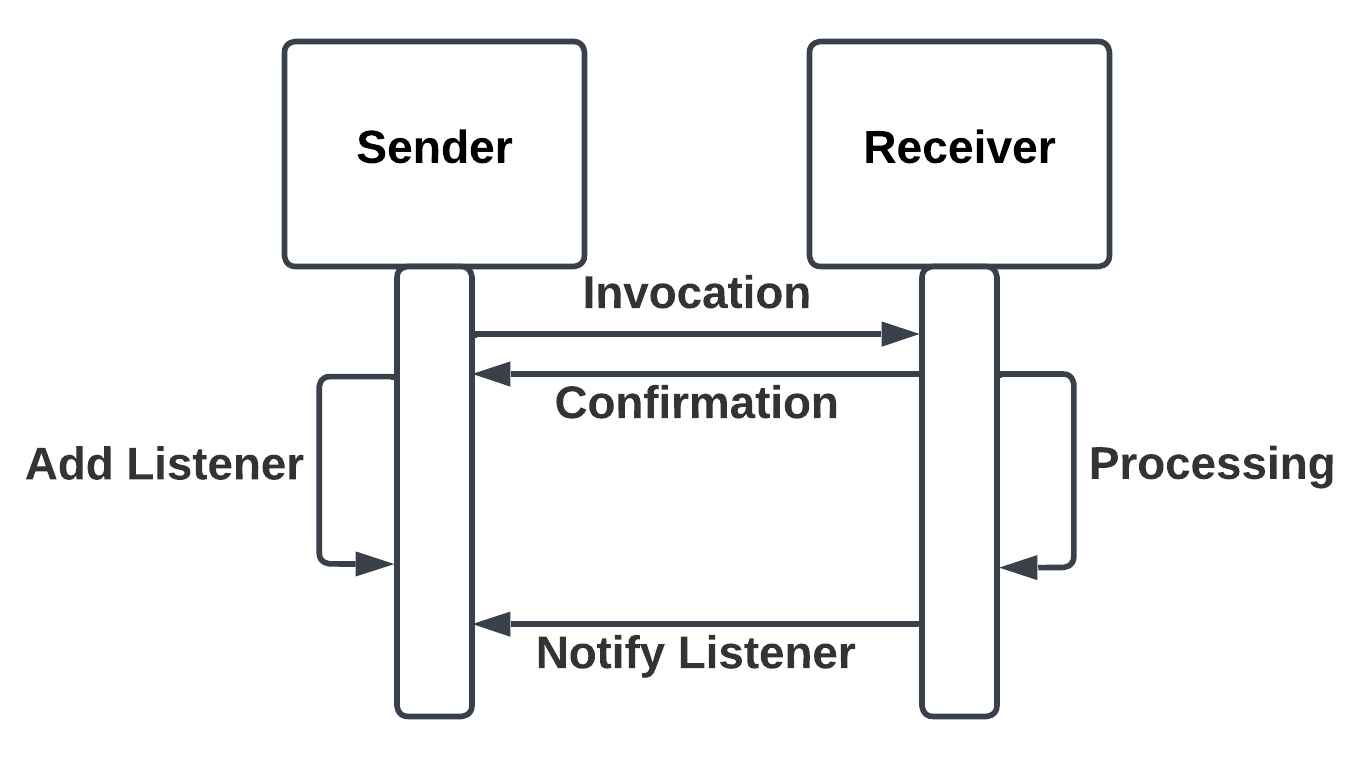}%
\label{asyn_call}}
\caption{Synchronous and Asynchronous Agent Communication Modes.}
\label{two_com_mode}
\end{figure*}

Fig. \ref{syn_call} and Fig. \ref{asyn_call} show the differences between synchronous and asynchronous communication modes. In synchronous mode, the sender agent will be blocked once it sends the invocation to the receiver. In this case, if the sender agent does not receive replies as expected, it is easy to get stuck and will not respond to other requests or perform behaviours. Since cloud computing is an open system that introduces various uncertainties, the synchronous communication mode can decrease the cloud QoS. In the asynchronous communication mode, the result listener will be added to the sender agent for monitoring and receiving replies from other agents. In this case, the sender agent will not be blocked during the communications and can perform other behaviours, which increases the robustness of the framework. 

\subsubsection{User Agent Definition}
The user agent module includes a set of user agents $UA=\left\lbrace ua_{1},..., ua_{n}  \right\rbrace$. Each user agent is responsible for one user. The n-th user agent is defined as $ua_{n}=\left\langle uid_{n}, B_{n},D_{n},I_{n}  \right\rangle$, where $uid_{n}$ is the agent ID, $B_{n}$ is the beliefs of $ua_{n}$, $D_{n}$ is the desires of $ua_{n}$, and $I_{n}$ is the intentions of $ua_{n}$. 

The belief of $ua_{n}$ is the information of the managed user, i.e. $B_{n}= \left\lbrace u_{n}  \right\rbrace$. The user agent regards the user task requirements and deadline as its belief and monitors their changes. Once the belief changes, the user agent will deliberate what goal should be achieved in the current situation and reason about how to achieve the goal.
Two top-level desires of user agents are (1) finding available VMs for tasks in the scheduling stage and (2) maximizing the task success rate in the rescheduling stage. User agents adopt different intentions to achieve these two desires under different situations. Specific intentions of three types of agents in both scheduling and rescheduling stages will be described in Subsections \ref{initial_scheduling} and \ref{rescheduling_sub}, respectively.

\subsubsection{Host Agent Definition}
The host agent module includes a set of host agents $HA=\left\lbrace ha_{1},..., ha_{i}  \right\rbrace$. Each host agent is responsible for one host. The i-th host agent is defined as $ha_{i}= \left\langle hid_{i}, B_{i}, D_{i}, I_{i} \right\rangle$, where $hid_{i}$ is the agent ID, $B_{i}$ is the agent beliefs, $D_{i}$ is the agent desires, and $I_{i}$ is the agent intentions.

The belief of $ha_{i}$ includes the managed host and its VMs information, where $B_{i}=\left\lbrace H_{i} \right\rbrace$. The host agent monitors the changes in VMs timely and triggers different desires and intentions. Two top-level desires of host agents are: (1) providing available time slots for received tasks in the scheduling stage and (2) maximizing the success rate of allocated tasks in the rescheduling stage. 

\subsubsection{Supervise Agent Definition}

The supervise agent module $SA=\left\lbrace sa \right\rbrace$ includes one supervise agent that is responsible for: (1) coordinating behaviours among user agents and host agents.
The supervise agent is defined as $sa=\left\langle sid, B, D, I \right\rangle$, $sid$ is the agent ID, $B$ is the agent beliefs, $D$ is the agent desires, and $I$ is the agent intentions.

The supervise agent synchronizes the information of VMs in hosts and treats the VMs information as beliefs, where $B=\left\lbrace H_{1},...,H_{i} \right\rbrace$. The top-level desire of the supervise is to: (1) recommend available VMs for user agents in both the scheduling and rescheduling stages; and (2) balance the resource utilization among VMs.

\begin{figure*}[h!]
\centering
\includegraphics[width=6in]{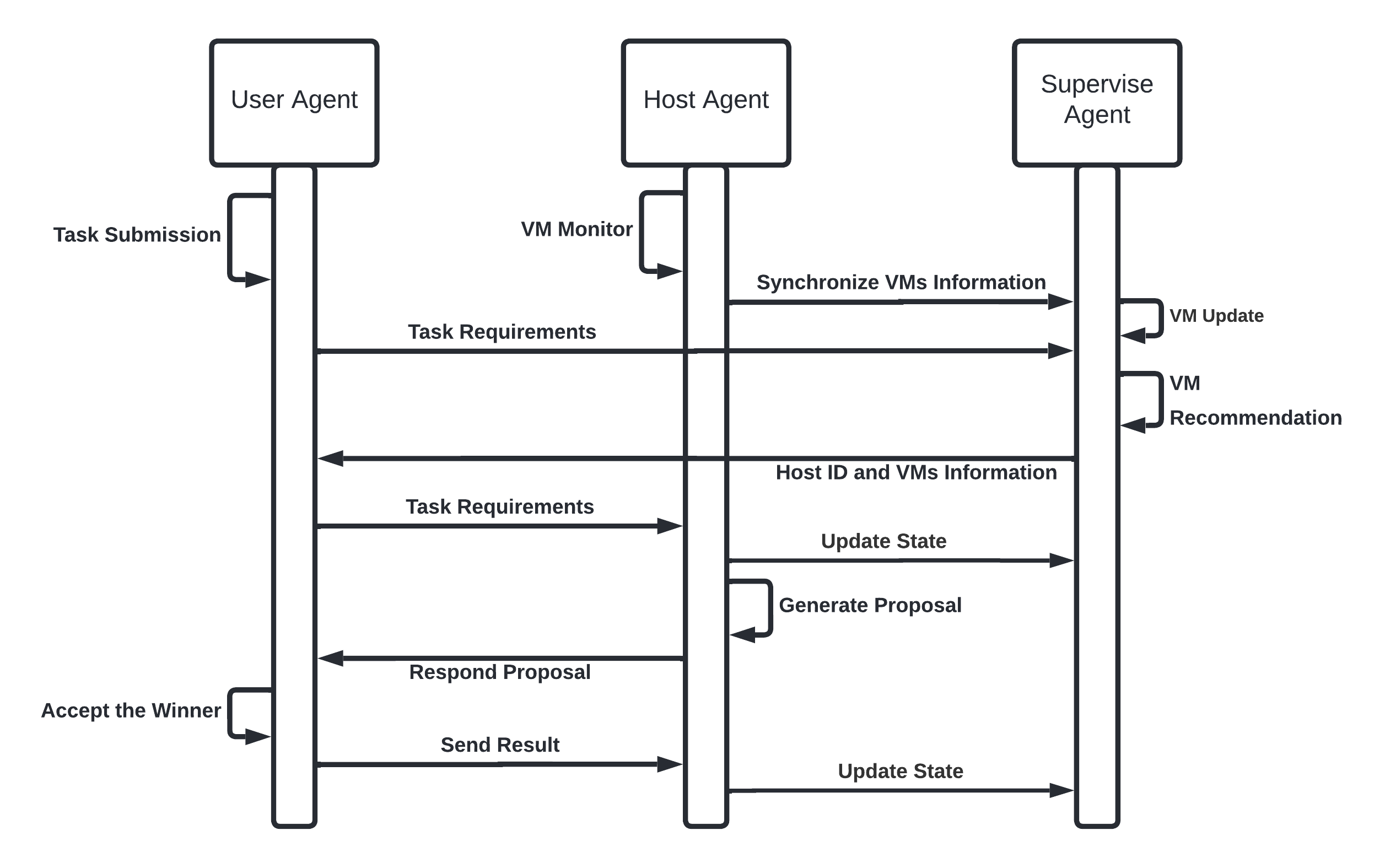}
\caption{Initial Scheduling Stage Sequence Diagram}
\label{seq_dig}
\end{figure*}

\subsection{Initial Scheduling}\label{initial_scheduling}

The \textbf{initial scheduling} stage refers to the period from task submission to task execution. In this stage, the primary responsibility of the three agent modules is to allocate computing resources for tasks and to achieve the two scheduling objectives, (1) minimizing the task makespan and (2) balancing the VMs resource utilization.

In this stage, three types of agents get information from users and hosts, and they implement the task-scheduling process through asynchronous communication and decision-making. Fig. \ref{seq_dig} shows the sequence diagram of this stage, and the procedures are described as follows.

\textbf{Step 1.} User agents get real-time task requirements from users through the user interface (UI); host agents get real-time VM information from hosts in the cloud center.

\textbf{Step 2.} Host agents share information about VMs with the supervise agent. Each time the available time of a VM changes, the corresponding host agent will synchronize the available time with the supervise agent. The supervise agent sorts VMs according to their available time $at_{i}^{k}(\tau)$. The earlier the $at_{i}^{k}(\tau)$, the higher the priority.

\textbf{Step 3.} When a user requires to use computing resources, the corresponding user agent will send the task requirements $T_{n}$ and the deadline $D_{n}$ to the supervise agent. The supervise agent recommends available VMs to the user agent through the Asynchronous Recommendation Algorithm (ARA). Then the user agent will receive a number of host agent ID $hid_{i}$ and VMs information.

\textbf{Step 4.} Each user agent initiates interactions with host agents through received $hid_{i}$ and request to use recommended VMs. 
The user agent sends the $T_{n}$, $D_{n}$ to host agents. Host agents will find the specified VMs and evaluate the task requirements, and then send proposals to the user agent. The proposal includes the start use time and expected completion time for the user. 

\textbf{Step 5.} If the user agent receives one or more proposals, it will accept the best proposal and reject others. In this paper, the user agent selects the best proposal based on the heuristic strategy named \textbf{Minimum Completion Time}, where the user agent tends to select the proposal that provides the earliest expected completion time \cite{nagadevi2013survey}. Otherwise, if the user agent does not receive any proposal from host agents, it will re-send requests to the supervise agent periodically until successful.

\textbf{Step 6.} After user agents and host agents reach agreements on resource allocation, the host agent will update the VMs information with the supervise agent.

\subsubsection{Asynchronous Recommendation Algorithm}

In the scheduling stage, the supervise agent $sa$ interacts with user agents and host agents and executes both the VMs update and VMs recommendation timely. However, the VMs update occurs after the user and host agents agree on resource allocation. The $sa$ can only realize whether the VM information will change once host agents inform it. During this period, the $sa$ will recommend the identical VM for user agents based on the previous VM information. The delay in VMs update can cause the $sa$ to recommend the same high-priority VMs to many users repeatedly. However, the available time of these VMs may have changed during agent interactions. This conflict will lead to an imbalance in VMs resource utilization and reduce the interaction efficiency between the user and host agents. 

To address the impact of VMs update delays, this paper proposes the Asynchronous Recommendation Algorithm (ARA). The proposed ARA coordinates two concurrent behaviours by labelling VMs with real-time states. Labelling states of VMs can avoid repeatedly recommending the same high-priority VMs to more than one user, so as to reduce the ineffective interactions between the user and host agents, as well as increase the scheduling efficiency.

Algorithm \ref{alg_1} describes the Asynchronous Recommendation Algorithm. The inputs of Algorithm \ref{alg_1} are users information and VMs in the datacenter. The outputs of Algorithm \ref{alg_1} are matches between users and VMs. 

\begin{algorithm}[h!]
 \caption{Asynchronous Recommendation Algorithm}
 \label{alg_1}
 \begin{algorithmic}[1]
 \renewcommand{\algorithmicrequire}{\textbf{Input:}}
 \renewcommand{\algorithmicensure}{\textbf{Output:}}
 \REQUIRE $U=\left\lbrace u_{1},...,u_{n} \right\rbrace$, $DC=\left\lbrace H_{1},..., H_{i} \right\rbrace$
 \ENSURE  match users and VMs
 \\ \textit{Initialisation} :$ua_{n}$ sets the n-th user information as its belief, $B_{n}=\left\lbrace u_{n} \right\rbrace$;
$ha_{i}$ sets VMs in $H_{i}$ as its belief, $B_{i}= \left\lbrace H_{i}=\left\lbrace vm_{i}^{1},..., vm_{i}^{k} \right\rbrace \right\rbrace$;
all VMs set states $state_{i}^{k}(\tau)=STATE\_READY$;
$sa$ sets VMs in hosts as its belief, $B = \left\lbrace H_{1},..., H_{i} \right\rbrace$;

\FORALL{$vm_{i}^{k}$ \label{vm_syn1}}
\IF{the information of $vm_{i}^{k}$ changes}
\STATE $ha_{i}$ triggers $i\in I_{i}$ to synchronize with $sa$;
\STATE $sa$ sorts VMs timely, the earlier the $at_{i}^{k}$, the higher the priority;\label{vm_syn2}
\ENDIF
\ENDFOR
\WHILE{$ua_{n}$ detects unallocated tasks \label{ua_send1}}
\STATE $ua_{n}$ triggers $d \in D_{n} $ for matching an appropriate VM;
\STATE $ua_{n}$ sends the $T_{n}$ and $D_{n}$ to $sa$;\label{ua_send2}
\STATE $sa$ searches for VMs that meets $T_{n}$, subject to: $ram_{i}^{k} \geq \max\limits_{T_{n}}( ram_{n}^{p})$ \& $sc_{i}^{k} \geq \max\limits_{T_{n}}(sc_{n}^{p})$ \& $bw_{i}^{k} \geq \max\limits_{T_{n}} (bw_{n}^{p})$ \& $at_{i}^{k}(\tau)+(\dfrac{\sum_{1}^{p} wl_{n}^{p}}{cpu_{i}^{k}}) \leq D_{n}$\; \label{req_1}
\REPEAT
\STATE $sa$ checks the state of each appropriate $vm_{i}^{k}$;\label{rep_1}
\IF{$state_{i}^{k}(\tau)==STATE\_READY$}
\STATE $sa$ adds $vm_{i}^{k}$ to the message $msg$;
\STATE set the state of $vm_{i}^{k}$ as $state_{i}^{k}(\tau)=STATE\_BUSY$; \label{rep_2}
\ENDIF
\UNTIL{the proposal size $\geq \theta \in \mathbb{N_+}$ $||$ no more VMs\; \label{end_con1}}
\STATE $sa$ sends the proposal to $ua_{n}$;\label{end_con2}
\IF{$ua_{n}$ does not receive a proposal \label{no_vm1}}
\STATE $ua_{n}$ sends the $T_{n}$ and $D_{n}$ to $sa$ periodically; \label{no_vm2}
\ELSIF{$ua_{n}$ reaches an agreement with a $ha_{i}$ \label{success_1}}
\STATE $ua_{n}$ and $ha_{i}$ synchronize information with $sa$;
\STATE set state of VMs as $state_{i}^{k}(\tau)=STATE\_READY$;\label{success_2}
\ENDIF
\ENDWHILE
 \end{algorithmic}
 \end{algorithm}
 
User agents and host agents get information from users and hosts, and treat the information of users and VMs as beliefs, respectively. Initially, all VMs states are set as ready, i.e., $STATE\_READY$. The supervise agent $sa$ stores the VMs information in hosts as its belief. Once the information of a VM $vm_{i}^{k}$ changes, the host agent $ha_{i}$ will synchronize the information changes with $sa$. Then the $sa$ will prioritize all the VMs based on the earliest available time. The earlier the available time, the higher the priority (Lines \ref{vm_syn1}-\ref{vm_syn2}).

When the user agent $ua_{n}$ detects unfinished tasks in $u_{n}$, it will trigger the desire to find available VMs for tasks. The $ua_{n}$ will send its task summary to $sa$, and the $sa$ will evaluate the task requirements $T_{n}$ and the specified deadline $D_{n}$ (Lines \ref{ua_send1}-\ref{ua_send2}). Then the $sa$ searches for appropriate VMs for $ua_{n}$ from high-priority to low-priority based on RAM, storage, bandwidth requirements, as well as the expected completion time cannot exceed $D_{n}$ (Line \ref{req_1}). Next, the $sa$ checks the states of each $vm_{i}^{k}$. If the state $state_{i}^{k}(\tau)$ at current time $\tau$ is ready, the $sa$ will add the $vm_{i}^{k}$ to the proposal. Then the state of $vm_{i}^{k}$ will be set as $STATE\_BUSY$ (Lines \ref{rep_1}-\ref{rep_2}). The $sa$ recommends $\theta$ number of VMs to a user per round. The value of $\theta$ affects the interaction efficiency, which will be discussed in the Experiment. Once the proposal includes $\theta$ number of VMs, or there is no more appropriate VMs at time $\tau$, the $sa$ will send the proposal to $ua_{n}$ (Lines \ref{end_con1}-\ref{end_con2}). If $ua_{n}$ does not receive any proposal, $ua_{n}$ will re-send its task summary to $sa$ periodically (Lines \ref{no_vm1}-\ref{no_vm2}). If the $ua_{n}$ receives a number of proposals and successfully matches a $vm_{i}^{k}$ for its tasks, $ua_{n}$ and $ha_{i}$ will update their information to $sa$, and the state of these VMs will be set as $STATE\_READY$ again (Lines \ref{success_1}-\ref{success_2}).

\subsection{Rescheduling}\label{rescheduling_sub}

After the initial task scheduling stage, tasks will wait for executions on VMs with specified time slots. From the waiting interval until task completion, uncertain events $E=\left\lbrace e_{1},..., e_{j} \right\rbrace$ may lead to tasks cannot be executed as pre-scheduled. In cloud computing environments, uncertain events such as VMs shut down, network latency, or improper operations by users, can cause tasks to execute longer than estimated times. Urgent tasks with strict deadlines may fail due to such reasons. In this case, rescheduling algorithms are required to reallocate resources for affected tasks, so as to improve the success rate of tasks.

Traditional rescheduling algorithms (e.g., predictive-reactive scheduling and periodical rescheduling algorithms) resolve uncertain events in a centralized manner. Because cloud computing is an open environment where streams of uncertain events will occur randomly, the stream of events will cause the central scheduler to perform a considerable amount of repetitive computations, so it is hard to find solutions for uncertain events because the following events may invalidate previous decisions. 

To fill this research gap, this paper proposes an agent-based rescheduling algorithm for cloud computing. Unlike traditional rescheduling algorithms, agent-based rescheduling algorithms can solve uncertain events case-by-case in a decentralized manner, therefore reducing the mutual influence between different events. Individual agents can get the feasible solution for uncertain events quickly, rather than repeatedly calculating all the feasible solutions to get an optimal schedule. Reducing the computation time of rescheduling for deadline-critical tasks can effectively improve the task success rate. 

In the rescheduling stage, user agents and host agents are responsible for monitoring the task executions on VMs and detecting the occurrence of uncertain events. BDI agents have the capability to real-time monitor the changes in their beliefs. Once the belief changes, BDI agents can trigger intentions to achieve desires. In the rescheduling stage, the desires of both user agents and host agents are to maximize the task success rate. 

Algorithm \ref{alg_2} describes the rescheduling algorithm in our framework. The inputs of Algorithm \ref{alg_2} are user information, VMs in the datacenter, and uncertain events. The outputs of Algorithm \ref{alg_2} are to resolve uncertain events for affected users. 

\begin{algorithm}[h!]
 \caption{Rescheduling Algorithm}
 \label{alg_2}
 \begin{algorithmic}[1]
 \renewcommand{\algorithmicrequire}{\textbf{Input:}}
 \renewcommand{\algorithmicensure}{\textbf{Output:}}
 \REQUIRE $U$, $DC$, $E=\left\lbrace e_{1},..., e_{j} \right\rbrace$
 \ENSURE  resolve uncertain events $E$
 

\IF{$\exists e_{j} \in u_{n}$ \label{user_change1}}

\STATE $ua_{n}$ detects the changes in $B_{n}$, $B_{n}=\left\lbrace u_{n} \rightarrow u_{n}^{'}=\left[ T_{n}^{'}, D_{n}^{'} \right] \right\rbrace$; \label{user_change2}
\IF{current schedule is invalid \label{trigger_1}}
\STATE $ua_{n}$ triggers $D_{n}$ for maximizing task success rate;
\STATE $ua_{n}$ deliberates intentions $I_{n}$, $I_{n}=\left[ i_{1} \preceq i_{2} \preceq i_{3} \right]$;\label{trigger_2}
\REPEAT
\STATE $ua_{n}$ processes intentions in priority;
\STATE \# Let $i_{1}$ denote changing time slots in the same $vm_{i}^{k}$; \label{intention_1}
\STATE \# Let $i_{2}$ denote changing a VM $vm_{i}^{k'}$ in the same host $H_{i}$;\label{intention_2}
\STATE \# Let $i_{3}$ denote changing a VM $vm_{i'}^{k'}$ in a different $H_{i'}$;\label{intention_3}

\UNTIL{resolve the $e_{j}$}
\ENDIF

\ELSIF{$\exists e_{j} \in vm_{i}^{k}$ \label{ha_tri1}}
\STATE $ha_{i}$ detects the changes in $vm_{i}^{k}$, $B_{i}=\left\lbrace vm_{i}^{k} \rightarrow \hat{vm}_{i}^{k} \right\rbrace$;
\STATE $ha_{i}$ triggers $D_{i}$ for maximizing task success rate;\label{ha_tri2}

\FORALL{$u_{n}$ on $\hat{vm}_{i}^{k}$ \label{search_h1}}

\IF{$\hat{vm}_{i}^{k}$ cannot satisfy  $T_{n} \| D_{n}$}
\STATE $ha_{i}$ searches for available VM for $u_{n}$ in $H_{i}$;\label{search_h2}

\IF{$\exists vm_{i}^{k*}$ satisfies $u_{n}$\label{ex_1}}
\STATE $ha_{i}$ generates proposal and informs $ua_{n}$;
\STATE $ua_{n}$ set the new contract with $ha_{i}$;\label{ex_2}
\ENDIF

\ELSIF{$\nexists vm_{i}^{k*}$ satisfies $u_{n}$ \label{noex_1}}
\STATE $ha_{i}$ informs $ua_{n}$;
\REPEAT
\STATE $ua_{n}$ interacts with $sa$ for new VM; \label{noex_2}
\UNTIL{resolve the $e_{j}$}
\ENDIF
\ENDFOR
\ENDIF
 \end{algorithmic}
 \end{algorithm}

User agents and host agents detect uncertain events $E$ in users and hosts, respectively. For each uncertain event $e_{j}$, if $e_{j}$ occurs in the user $u_{n}$, the user agent $ua_{n}$ will recognize the information changes in user information (Lines \ref{user_change1}-\ref{user_change2}). The new user information $u_{n}^{'}=\left[ T_{n}^{'}, D_{n}^{'} \right]$ includes the new task requirements $T_{n}^{'}$ and new deadline $D_{n}^{'}$. The user agent $ua_{n}$ evaluates the new task requirements and deadline, and checks if the current $vm_{i}^{k}$ can meet the new $T_{n}^{'}$ or $D_{n}^{'}$. If the current $vm_{i}^{k}$ cannot meet the new user information, user agent $ua_{n}$ will trigger the desire for maximizing task success rate and deliberate intentions (Lines \ref{trigger_1}-\ref{trigger_2}). 

This paper considers three intentions in priority order. 
The highest priority intention $i_{1}$ is to renegotiate time slots with the current $vm_{i}^{k}$ (Line \ref{intention_1}). In this case, $ua_{n}$ can interact with the host agent $ha_{i}$ directly. If $vm_{i}^{k}$ can meet the $T_{n}^{'}$ and can also provide new contract within the $D_{n}^{'}$, then $ha_{i}$ will provide a new contract with new time slots to $ua_{n}$.
The second priority intention $i_{2}$ means that, if the current $vm_{i}^{k}$ cannot meet the new user information, the host agent $ha_{i}$ will search for available VMs in host $H_{i}$ (Line \ref{intention_2}). If there is a new VM $vm_{i}^{k'}$ in host $H_{i}$ can meet $u_{n}^{'}$, the host agent $ha_{i}$ will generate and reply the new proposal to $ua_{n}$. Then $ua_{n}$ will accept the new proposal and set a new contract with the $ha_{i}$.
The lowest priority intention $i_{3}$ is to re-interact with $sa$ for VM recommendations if intentions 1 and 2 do not work (Line \ref{intention_3}). 

To reduce the impact of uncertain events on other task executions, the range of three intentions is from local rescheduling to global rescheduling. Intention $i_{1}$ limits the impact of uncertain events within the same virtual machine, while $i_{2}$ limits the impact within the same host. These two high-priority intentions can effectively reduce agent communications during the rescheduling stage, so as to improve the rescheduling efficiency. If these two intentions cannot solve the uncertain event $e_{j}$, $ua_{n}$ will process the $i_{3}$ which re-processes the initial scheduling stage. The $ua_{n}$ will repeat these three intentions until the $e_{j}$ is solved. Because at different runtimes, the external environments of agents are different. Periodically repeating these three intentions can avoid the impact of information delay, thereby improving the task success rate. 

If the uncertain event $e_{j}$ exists in the VM $vm_{i}^{k}$, the host agent $ha_{i}$ will detect the changes within its belief $B_{i}$ and trigger the desire for maximizing task success rate (Lines \ref{ha_tri1}-\ref{ha_tri2}). Then, for each user $u_{n}$ that has been assigned on the $vm_{i}^{k}$, $ha_{i}$ will check if the current attributes of $vm_{i}^{k}$ can satisfy the user information. If the current VM $\hat{vm}_{i}^{k}$ cannot satisfy the user task requirements $T_{n}$ or deadline $D_{n}$, $ha_{i}$ will firstly search for available VMs for the $u_{n}$ (Lines \ref{search_h1}-\ref{search_h2}). If there is an available $vm_{i}^{k*}$ in $H_{i}$ that can meet the $u_{n}$, $ha_{i}$ will generate a new proposal for $u_{n}$ and send the proposal to $ua_{n}$. Then $ua_{n}$ will accept the new proposal and set a new contract with $ha_{i}$ (Lines \ref{ex_1}-\ref{ex_2}). In this case, the impact of $e_{j}$ can be limited within the same host. Otherwise, if there is no one VM that can meet the user information, $ha_{i}$ will inform the $ua_{n}$. The $ua_{n}$ will re-interact with $sa$ for new VM recommendations until match a new VM for $u_{n}$ (Lines \ref{noex_1}-\ref{noex_2}).

\section{Experiment}\label{experiment}

In the experiment, the proposed task scheduling framework was implemented in Java using JADEX, and the framework was tested on CloudSim. We conducted two experiments to evaluate the proposed framework. \textbf{Experiment 1} is to evaluate the initial scheduling process with the proposed asynchronous recommendation algorithm. In this stage, two scheduling objectives are to minimize the task makespan and to balance the resource utilization among virtual machines. \textbf{Experiment 2} is to test the rescheduling process with the occurrence of uncertain events. In this stage, the objective is to maximize the success rate of tasks.

\subsection{Settings and Results of Experiment 1}

\begin{table}[t]
\caption{Experiment 1 settings}
\setlength{\tabcolsep}{11mm}
 \label{E1_setting}
\begin{tabular}{c|c}
\hline
\textit{Parameters}&\textit{Values}\\
\hline
Number of Hosts& [10, 50]\\
\hline
Number of Users & 10000\\
\hline
Number of VMs in a Host & [10, 20]\\
\hline
Number of Tasks in a User & [5, 10]\\
\hline
\end{tabular}
\end{table}

\begin{table}[t]
\caption{Settings of VMs in Experiment 1}
\setlength{\tabcolsep}{12mm}
 \label{E1_VM}
\begin{tabular}{c|c}
\hline
\textit{Parameters}&\textit{Values}\\
\hline
CPU (MIPS) & [500,2500]\\
\hline
RAM (M) & [1250, 1740]\\
\hline
Storage (G) & [4, 10]\\
\hline
Bandwidth (MB/s) & [1000, 2000]\\
\hline
\end{tabular}
\end{table}

\begin{table}[t]
\caption{Settings of Tasks in Experiment 1}
\setlength{\tabcolsep}{11mm}
 \label{E1_task}
\begin{tabular}{c|c}
\hline
\textit{Parameters} &\textit{Values}\\
\hline
Workload (MI)&[10000, 40000]\\
\hline
RAM (M) & [800, 1200]\\
\hline
Storage (G) & [1, 8]\\
\hline
Bandwidth (MB/s) & [100, 500]\\
\hline
\end{tabular}
\end{table}

Parameters used in Experiment 1 are shown in \textbf{Table \ref{E1_setting}}. The detailed settings of VMs and tasks are shown in \textbf{Table \ref{E1_VM}} and \textbf{\ref{E1_task}}, respectively. In Experiment 1, because the first objective is to evaluate the makespan, so we set the deadline $D_{n}$ of each user large enough to ensure the task success rate to be $100\%$. The parameter $D_{n}$ will be considered and discussed in Experiment 2.

\begin{figure}[h!]
\centering
\includegraphics[width=3.3in]{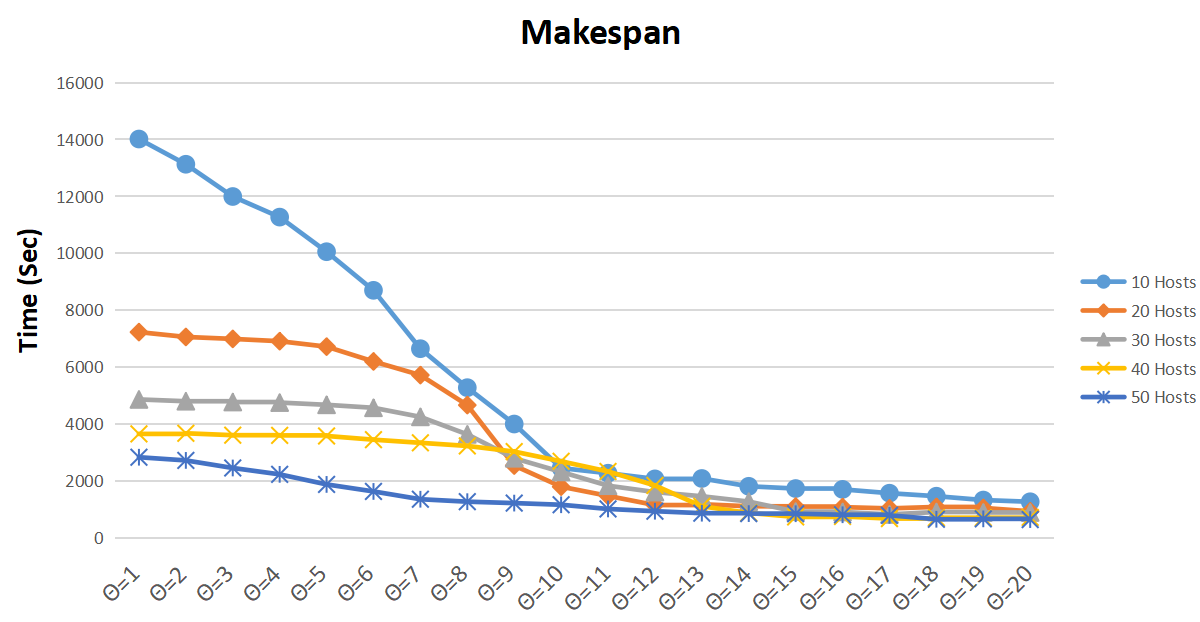}
\caption{Makespan of our framework}
\label{makespan}
\end{figure}

\begin{figure}[h!]
\centering
\includegraphics[width=3.3in]{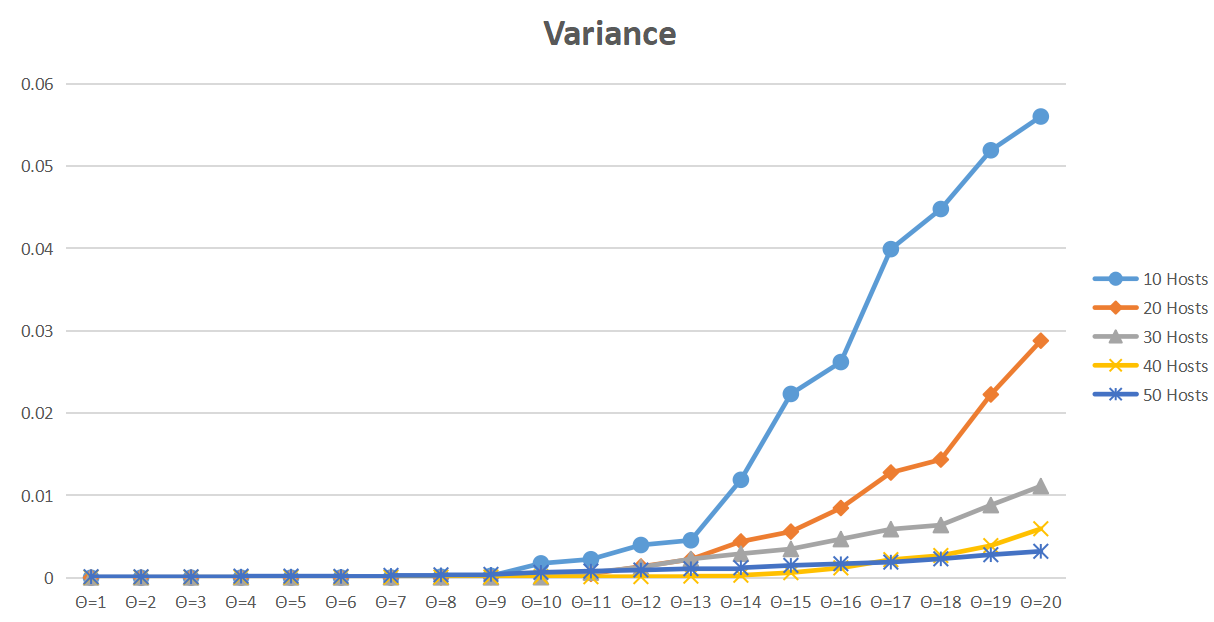}
\caption{Variance $\mathbb{V(\mu)}$ of our framework}
\label{variance}
\end{figure}

Fig. \ref{makespan} and Fig. \ref{variance} show the evaluation of our proposed framework under different experimental settings. Fig. \ref{makespan} shows the makespan for 10000 users, where the $\theta$ value was set between $\left[ 1, 20 \right]$ and the number of hosts was set between $\left[ 10, 50 \right]$. The $\theta$ value represents the number of VMs that the supervise agent recommends each round to user agents. The number of hosts represents the degree of resource scarcity, and the increment represents the resource going from scarcity to abundance. Fig. \ref{variance} shows the resource utilization variance under the same settings.

Fig. \ref{makespan} shows that the task makespan will decrease as the $\theta$ value increases. Because when the $\theta$ is small, resource allocation is mainly determined by the supervise agent, whose main focus is to balance the resource utilization among VMs. In this case, user agents receive only a few VM recommendations per round from the supervise agent, so it is difficult to optimize the task makespan. As $\theta$ increases, user agents receive and consider more proposals from host agents when making decisions. Powerful VMs can provide optimized completion time, thus reducing the overall makespan. 

Fig. \ref{makespan} also shows that when resources are abundant (i.e., hosts number = 50), even though the $\theta$ is small (from 1 to 5), the makespan will not exceed 3000. With the increase of $\theta$, the decrease of makespan is not large, and it gradually stabilizes after $\theta$ is greater than 6. When the computing resources are not enough (i.e., hosts number = 10), small $\theta$ value results in a large makespan. This is caused by the lack of resources and excessive pursuit of resource utilization balancing. As $\theta$ increases, the makespan decreases rapidly and becomes stable when $\theta$ is greater than 13. Because when the total number VMs is not large, the increase of $\theta$ will enable user agents to consider a large percentage of VMs and to interact with most of host agents, so as to reach an approximate optimal solution.

The increase in $\theta$ can lead to an imbalance in resource utilization. As Fig. \ref{variance} shows, when the $\theta \in \left[1, 9 \right]$, the variance $\mathbb{V(\mu)}$ is quite small. In this case, the supervise agent plays an important role in balancing the utilization among VMs. With the increase of $\theta$, the influence of supervise agent on resource allocation gradually decreases. User agents and host agents are mainly focusing on the minimization of makespan, so as to increase the $\mathbb{V(\mu)}$.

\begin{table}[t]
\caption{Comparison with other methods}
\setlength{\tabcolsep}{5.5mm}
 \label{comparison}
\begin{tabular}{c|c|c}
\hline
\textit{Methods}& Makespan (sec) & Variance $\mathbb{V(\mu)}$ \\
\hline
MCT &2396 & 4.236e-3\\

MET& 9364 & 1.064e-1\\
Min-Min & 2743 & 7.528e-2\\
Round Robin&5914 & 2.835e-2 \\
Gupta et al. \cite{gupta2017load}&7856 & 5.992e-5\\
Yao et al. \cite{yao2017cloud}&2352&7.952e-2\\
Singh et al. \cite{singh2015autonomous}&3806 &8.954e-6\\

\hline
\end{tabular}
\end{table}

We conducted the comparative tests with seven existing methods under the same VMs settings with 20 hosts and 10000 users. The selected seven methods include two static scheduling methods MCT and MET, two dynamic scheduling methods, Min-Min and Round-Robin \cite{priyadarsini2014performance}, two non-agent-based scheduling methods \cite{gupta2017load,yao2017cloud} and one agent-based scheduling model. The comparison results show in Table \ref{comparison}.

Comparing two static methods, the MCT achieved a small makespan because it takes the task completion time as the main consideration. Due to the main calculation in MET being the task execution time, when the VMs capability differences are large, MET prefers to allocate tasks on powerful VMs, resulting in a high makespan and imbalance of resource utilization. As for the two dynamic scheduling methods, the Min-Min can reach a near-optimal makespan with imbalance resource utilization, while the Round Robin can decrease the variance with a higher makespan. 
The scheduling model proposed by Gupta et al. \cite{gupta2017load} focuses on load balancing, which performs well on variance. The model proposed by Yao et al. \cite{yao2017cloud} had a minimized makespan because this model employs an improved genetic algorithm to optimize the schedule, so as to minimize the task makespan. The agent-based model \cite{singh2015autonomous} proposed the load agent to balance the utilization among VMs, effectively reducing the variance, while the makespan is not large. 

\subsubsection{Discussion}

The experimental results suggest a trade-off between the task makespan and resource utilization balancing. Most of the existing methods only focus on one of the scheduling objectives (i.e., makespan or utilization balancing). Our proposed agent-based scheduling framework can adapt to these two scheduling objectives by adjusting the value of $\theta$. Specifically, the increase of $\theta$ can minimize the task makespan, while increasing the variance. Once the $\theta$ increases, the decision-making and communications between user agents and host agents increase correspondingly. Besides, when the ratio of resources to tasks changes, the optimal $\theta$ also changes. Hence, in actual use, the $\theta$ value should be dynamically adjusted according to the differences in the ratio of tasks and resources, as well as considering the number of users and their tasks, so as to achieve an optimal scheduling performance.

\subsection{Settings and Results of Experiment 2}

In Experiment 2, the settings of VMs and tasks were the same as in Experiment 1. The number of users was set as 10000, and the number of hosts was set as 30. Based on the results of Experiment 1, the deadlines of users were set randomly between $\left[2000, 5000 \right]$, and the $\theta=5$. During the task execution, we conducted the series of uncertain events $E=\left\lbrace e_{1},..., e_{j} \right\rbrace$ to randomly occur in VMs and tasks. We assumed that each task or VM could only involve one uncertain event. Table \ref{event_setting} shows the settings of uncertain events in Experiment 2. 

\begin{table}[t]
\caption{Uncertain events $E$ settings}
\setlength{\tabcolsep}{1.5mm}
 \label{event_setting}
\begin{tabular}{c|c}
\hline
\textit{Parameters}& \textit{Values}\\
\hline
Task $wl$ $\&$ $ram$ $\&$ $sc$ $\&$ $bw$ & randomly increased by $\left[10\%, 50\% \right]$\\
\hline
User deadline $D$ & randomly decreased by $\left[ 100, 1000 \right]$ sec\\
\hline
VM $cpu$ $\&$ $ram$ $\&$ $sc$ $\&$ $bw$ &randomly decreased by $\left[ 10\%, 50\% \right]$\\
\hline
\end{tabular}
\end{table}

\begin{figure}[h!]
\centering
\includegraphics[width=3.3in]{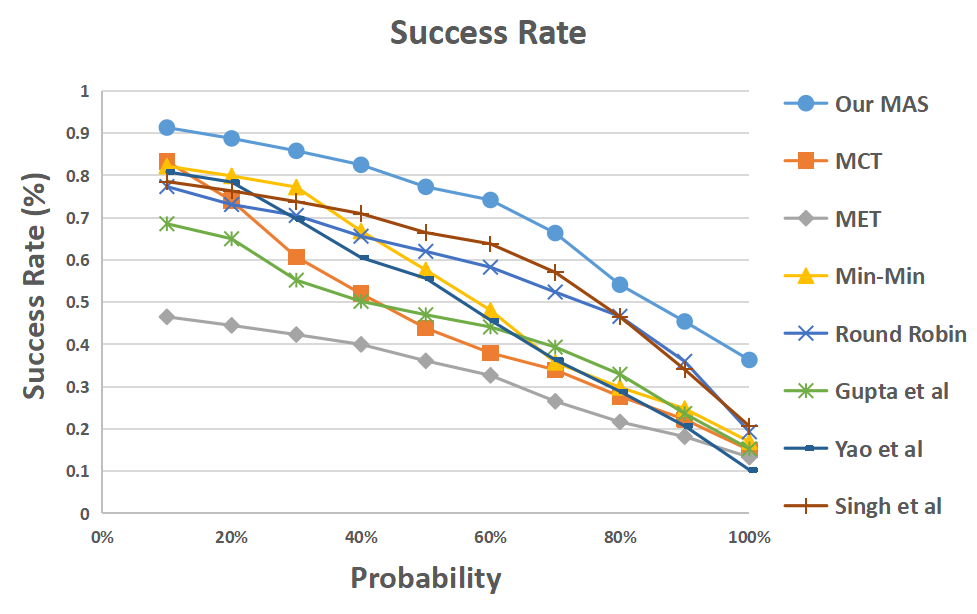}
\caption{Success rate of Experiment 2}
\label{success_rate_fig}
\end{figure}

Fig. \ref{success_rate_fig} shows the success rate of our proposed framework and seven comparison models under the same settings. When the uncertain events occurrence probability increased from $10\%$ to $100\%$, our proposed framework could reach the highest success rate for rescheduling (from $38\%$ to $91\%$). 
Because there is no rescheduling mechanism for the seven comparison models, these seven methods will reallocate resources for affected tasks when uncertain events occur. 

The two static methods MCT and MET have very different rescheduling performances. When the probability is small (from $10\%$ to $40\%$), the success rate of MCT is up to $82\%$. However, the success rate of MET belongs to $\left[40\%, 47\% \right]$. This is due to the large makespan introduced in the MET method. Because the VMs are heterogeneous with different CPU capabilities, the execution times of tasks on various VMs are different. MET tends to allocate tasks to the powerful VMs, thereby tasks with strict deadlines cannot be completed in time. As the probability increases, the success rates of the two methods decrease dramatically, because these two static methods involve a high computation time. Once uncertain events occur, especially when the VMs capability changes, the central scheduler needs to calculate the execution time or completion of each affected task on each VM. The upcoming events will also invalidate the previous rescheduling decisions, which is time-consuming. The high computation time decreases the task success rate, which also suggests that static methods are not flexible and robust in online environments.

As for the two dynamic methods, Min-Min and Round Robin, they treat the changed tasks as new arrival tasks which can be allocated to resources in the online mode. Lower response time and computations are involved, so these two dynamic methods have better success rates than that of static methods. 
The model proposed by Gupta et al. mainly focuses on load balancing while the makespan is ignored, so the success rate is lower than other methods when the probability is small. The method presented by Yao et al. applies the improved genetic method to optimize the makespan, which can reach a high success rate when the probability is small. However, when the probability increases, the high amount of computations increases the rescheduling response time, thereby decreasing the success rate significantly. The agent-based model proposed by Singh et al. has a better performance than other comparison methods because the rescheduling decisions were made by independent agents, which effectively decreases the response time through parallel computations.

The results illustrate that our rescheduling algorithm can effectively increase the task success rate. There are three main reasons. (1) Each uncertain event can be detected and resolved independently by the responsible BDI agents, which decreases the mutual influence between different uncertain events. BDI agents can automatically detect changes in their beliefs. Once the attributes of tasks or VMs change, BDI agents will trigger the rescheduling desire and corresponding intentions to reach the desired status, i.e., to maximize the task success rate. (2) Our framework prioritizes resolving uncertain events in the local range, i.e., changing time slots for tasks in the same virtual machine or finding a new available virtual machine in the same host. Although solving uncertain events in the local range cannot achieve global optimality, it can quickly find feasible solutions, so as to reduce the rescheduling response time. For urgent tasks, reducing the rescheduling response time can effectively increase the rescheduling success rate. (3) Our framework iteratively resolves uncertain events until they are solved. At different runtimes, the external environments are different. Repeatedly performing the rescheduling processes allows more tasks to be allocated to computing resources, so as to increase the success rate. Since the amount of computation is distributed among independent agents, the computational complexity for repetitive executions is not considerable. 

\subsubsection{Discussion}

Experiment 2 suggests that the rescheduling performance is mainly determined by two factors: (1) task makespan and (2) rescheduling response time. Methods that can achieve optimized makespan have advantages in rescheduling because they can complete more tasks within limited deadlines. Besides, whether in the scheduling or rescheduling process, long response times may cause urgent tasks with strict deadlines cannot to be completed in time. In our framework, through the global recommendation of the supervise agent and the interactions between host and user agents, our framework can quickly find the near-optimal schedule for all involved tasks. In addition, our proposed framework effectively distributes the computational complexity through parallel computations among independent BDI agents, thereby reducing the response time and the mutual influence between the series of uncertain events. On the other hand, uncertain event detection is a significant problem in rescheduling. In the simulation stage, we conducted uncertain events for both non-agent-based and agent-based methods. However, in real applications, non-agent-based methods are hard to detect and collect uncertain events automatically in open and distributed environments. BDI agents have the advantage of detecting environmental changes and resolving uncertain events in distributed environments, which is in line with that of cloud computing.  

\section{Conclusion}\label{conclusion}

This paper investigated the task scheduling and rescheduling problems in cloud computing environments and proposed a decentralized agent-based framework for improving the scheduling and rescheduling performance. In this paper, we applied BDI agents with asynchronous communication mode to construct the framework, which has better robustness and efficiency than traditional methods. Besides, this paper proposed a novel asynchronous recommendation algorithm for the initial task scheduling stage and proposed an agent-based rescheduling algorithm that can limit the impact of uncertain events within the local range. The experimental results show that our proposed framework has advantages in minimizing the task makespan, balancing resource utilization, and maximizing the task success rate when typical uncertain events occur. In our future work, we will take realistic constraints in cloud computing and further optimize the behaviours of BDI agents to improve the scheduling and rescheduling performance.

\ifCLASSOPTIONcaptionsoff
  \newpage
\fi



\bibliographystyle{IEEEtran}
\bibliography{IEEE_trans.bib}
\end{document}